\documentclass{aa}

\usepackage{times}
\usepackage{psfig}
\usepackage{graphics}

\newcommand{\Edot}{$\dot E$}

\newcommand{\APJ}{ApJ\, } 
\newcommand{\ApJ}{ApJ\, }

\newcommand{\Nat}{Nat\, }
\newcommand{\Aa}{A\&A\,}

\newcommand{\MNRAS}{MNRAS\, }

\newcommand{\flux}{flux$^b$ $\!\!\times 10^{-13}$}
\newcommand{\Nh}{Nh $\!\times10^{21}$}

\newcommand{\Rate}{Rate$^a\times 10^{-3}$}

\begin{document}

   \thesaurus{06(                                         
               08.16.7: PSR B1055-52, B0656+14, Geminga; 
               08.14.1;                                  
               13.25.3)
              }

\title{The putative pulsar-wind nebulae of the three Musketeers PSR B1055-52, 
B0656+14 and Geminga revisited}

\author{W.~Becker\inst{1} \and N.~Kawai\inst{2} \and W.~Brinkmann\inst{1} \and R.~Mignani\inst{3}}  
\offprints{W.~Becker}
\mail{web@mpe.mpg.de}
\institute{Max-Planck-Institut f\"ur extraterrestrische Physik, D-85740 Garching bei M\"unchen, Germany \and
Cosmic Radiation Laboratory, RIKEN, 2-1 Hirosawa, Wako-shi, Saitama 351-01, Japan \and
ESA-STECF, D-85740 Garching bei M\"unchen, Germany}

\date{Received: May 20, 1999 / Accepted: Aug 25 1999}

\authorrunning{W.~Becker et al. }
\titlerunning{The putative pulsar-wind nebulae of the three Musketeers}
\maketitle

 \begin{abstract}  We report on the  analysis  of archival ASCA, ROSAT
 and  BeppoSAX  data of  PSR   B1055$-$52, PSR   B0656+14  and Geminga,
 performed in order  to investigate the  reality of the putative  $\sim
 10-20$  arcmin  wide X-ray pulsar-wind   nebulae  recently proposed to
 exist around these  pulsars. In all these  cases,  the ROSAT and/or
 BeppoSAX data  show that the diffuse  and clumpy X-ray nebulae found
 by ASCA  can be resolved  in the  contribution  of unidentified  point
 sources, unlikely related to the pulsars. 
 
 \end{abstract}

\section{Introduction}
 
  Rotation-powered pulsars are rapidly  spinning and strongly magnetized
  neutron stars  which are radiating  at the expense of their rotational
  energy.  This  interpretation, first  expressed   by Pacini (1967) few
  months before the discovery of radio  pulsars, is the basic message of
  the magnetic  braking model. The   model predicts that  the radiative 
  energy loss of a  magnetic dipole, co-rotating  with a neutron star at  
  a period $P$ and period derivative $\dot{P}$, corresponds to a decrease 
  of  the neutron star's rotational energy $\dot E \propto P^{-3}\dot{P}$. 
  Although   the magnetic braking  model   in its general meaning  is
  widely  accepted, the {\em  observed} spin-modulated emission, which
  gave  pulsars  their name,  is found   to account  only for  a small
  fraction of \Edot. Efficiencies $\eta = L/{\dot  E}$ observed in the
  radio and  optical bands are typically in  the range $\sim 10^{-7} -
  10^{-5}$  whereas  at  X- and   gamma-ray  energies they   are about
  $10^{-4} - 10^{-3}$ and $\sim 10^{-2} - 10^{-1}$, respectively. It is
  therefore  a long standing question how rotation-powered pulsars loose
  the bulk of their rotational energy. 

  That  the energy loss  of rotation-powered pulsars  can not be fully
  accounted for by  dipole  radiation is known from  investigation of
  the pulsars' braking index $n=2-P{\ddot P}  {\dot P}^{-2}$.  Pure dipole
  radiation  would imply  a braking index  $n=3$,   whereas the values
  observed  so  far   are $n=2.515 \pm   0.005$  for  the  Crab (Lyne
  et al.~1988), $n=2.8 \pm 0.2$  for PSR B1509$-$58 (Kaspi et al.~1994),  
  $n=2.28\pm 0.02$ for PSR B0540$-$69 (Boyd et al.~1995) and $n=1.4 \pm  
  0.2$  for the Vela  Pulsar  (Lyne et  al.~1996). The deviation from 
  $n=3$ is usually taken as evidence that a significant
  fraction of   the pulsar's  rotational energy is   carried off  by a
  pulsar wind i.e. a mixture of charged particles and electromagnetic
  fields, which,  if  the conditions are  appropriate,  forms a
  pulsar-wind   nebula  observable  at   optical,  radio-   and  X-ray
  energies. 

  Such  pulsar-wind nebulae  (often   called plerions  or  synchrotron
  nebulae)  are known so far  {\em only} for  few young and powerful
  (high $\dot E$)  rotation-powered pulsars and for some center-filled
  supernova remnants,  in which a  young neutron star  is expected but
  only emission from its plerion is  detected. The physical details of
  how the pulsar-wind is generated and how it  interacts with the ambient
  medium is only poorly understood. 
 
  The  existence  of X-ray bright
  pulsar-wind nebulae as a  common feature surrounding rotation-powered
  pulsars was recently   reported on the  basis   of ASCA observations
  (Kawai \& Tamura 1996).   Indeed,  the  ASCA    data  of    PSR
  B1055$-$52, B0656+14,   and Geminga (as well  as  of several other
  pulsars) show   a faint diffuse and   somewhat clumpy  emission on
  scales of $\sim 10-20$ arcmin around these objects, which has
  been interpreted
  as a result of the interaction between the pulsar wind outflow and the
  surrounding  interstellar  matter (Kawai  et  al.~1998a, 1998b; Shibata  
  et al.~1997). 
                  
  In this paper, we report on a joint analysis of the archival ASCA and
  ROSAT data   of the fields surrounding  PSR  B1055$-$52, B0656+14, and
  Geminga.   The aim of  the analysis  was  to  search  for soft X-ray
  emission from the putative ASCA detected  pulsar-wind nebulae and to
  check the possible  presence of X-ray  point sources in the pulsars'
  neighborhood, which -- due to the wide ASCA point-spread function of
  ($\simeq 3$ arcmin   FWHM) -- could  result in  patterns of  diffuse
  emission and knots.  For PSR  B1055$-$52 we additionally made use of
  X-ray  data  taken by  the  Narrow  Field  Instruments (NFI)  aboard
  BeppoSAX.  The objects were selected  according to the similarity in
  their spin-parameters  and X-ray emission  characteristics ({\it the
  three   musketeers},  see Becker    \&   Tr\"umper 1997). They   are
  furthermore the only sources with  ASCA detected pulsar-wind nebulae
  for which sufficiently deep ROSAT and/or BeppoSAX data exist.

\section{Data analysis and results}

\subsection{PSR B1055$-$52}
  The 197  ms pulsar  PSR  1055$-$52 is  a   middle aged neutron  star
  ($\tau=P/2\dot{P}=5.3  \times  10^5$  yr) with   a  magnetic  dipole
  component $B_\perp =  10^{12}$ G  and  a spin-down energy  of  $\log
  \dot{E}=34.48$ erg/s. 
  X-ray emission from the pulsar was first detected by Chen \& Helfand
  (1983) using the {\em Einstein} Observatory and subsequently by {\em
  EXOSAT} (Brinkmann \& \"Ogelman 1987) and ROSAT (\"Ogelman \& Finley
  1993).  The ROSAT data provided enough information to detect for the
  first  time the X-ray  pulses and to  discriminate between different
  X-ray emission  mechanisms (\"Ogelman \& Finley   1993 and Becker \&
  Tr\"umper 1997 for a summary).  

  ASCA observations of PSR B1055$-$52 were  performed in January 1995 
  with  the  Gas Imaging Spectrometers (GIS) and  Solid state Imaging  
  Spectrometers (SIS) detectors for an exposure time of 37 ksec and 34 
  ksec,  respectively.  Results on  a
  combined  spectral analysis of  the  pulsar X-ray emission based  on
  ROSAT   and  ASCA  data have  been    reported  by  Greiveldinger et
  al.~(1996).  Wang et al.~(1998)   confirmed later on that   the ASCA
  data  beyond 1   keV  are  best  fitted by   a power-law spectrum,
  associated with   non-thermal  emission  from  the  pulsar's   inner
  magnetosphere. 
 
\begin{figure}[h]
  \vspace{1ex}
   \centerline{\psfig{file=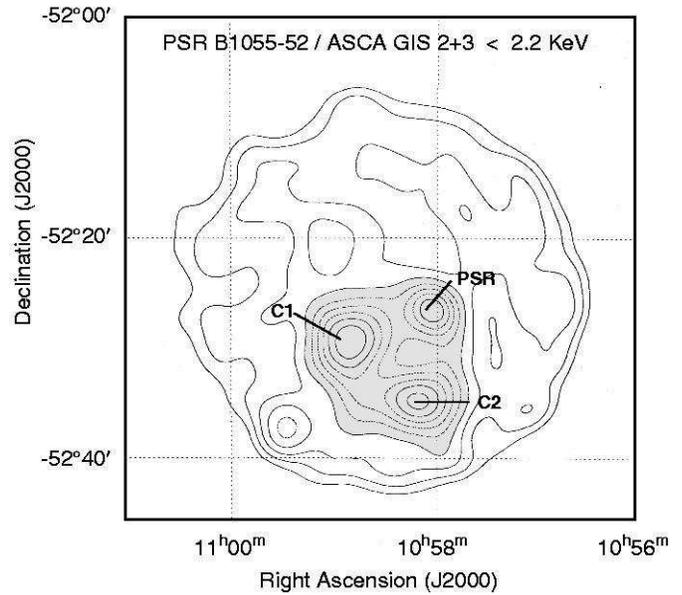,width=8.8cm,clip=}}
   \caption{Contour plot of the region around PSR B1055$-$52 as seen with the 
    ASCA GIS(2+3). Shown is the GIS full field of view for photon energies 
    below 2.2 keV. The image has been smoothed with a Gaussian filter to 
    increase the visibility of faint emission. The gray-shaded area indicates 
    the emission which  is proposed  to be from a pulsar-wind nebula  
    (Shibata et al.~1997) . \label{1055_ASCA}}
 \end{figure}

  The detection  of  a putative  pulsar-wind  nebula surrounding PSR
  B1055$-$52   was communicated   by   Shibata  et al.~(1997).  
  Fig.~\ref{1055_ASCA} shows a contour plot of the ASCA GIS (unit 2+3) full
  field  of view,  showing  the  position of  PSR  B1055$-$52  and its
  surroundings (gray shaded). The pulsar as well as two sources C1 and
  C2 are  indicated.  C1 and   C2   were interpreted by   Shibata   et
  al.~(1997) as  "clumps"  of X-ray   emission caused   by   an
  interaction of a pulsar-wind   outflow with the  local  interstellar
  environment. The   very low signal-to-noise  ratio   of the "clumps"
  limits the spectral analysis, resulting  in large uncertainties  in
  the best-fit parameters   and making it impossible   to discriminate
  between    different emission models   like,  e.g.,  a Raymond-Smith
  thermal   plasma or a power-law  (see  Table 1  and 2  of Shibata et
  al.~1997    for details), which   is  expected in  case of plerionic
  emission. 
 
  In order to cross-check the  ASCA results on  PSR B1055$-$52 and its
  surroundings, we  have  made use  of data  taken by   the Narrow Field
  Instruments (NFI)  aboard BeppoSAX.  Of  particular interest are the
  data  taken with  the    Medium Energy  Concentrator  (MECS),  which
  provides  a  slightly better   angular  resolution\footnote{The MECS
  on-axis point-spread  function has a  half energy width of about 1.7
  arcmin at  1.5 keV.} than  the  ASCA GIS, whereas  both  instruments
  cover a similar   energy  range. MECS  (unit 1-3)  has  observed the
  pulsar on 28$-$29th December 1996  for an effective exposure time of
  $\sim  56$ ksec.  The Low  Energy  Concentrator (LECS), which has an
  observing efficiency of only $\sim 40$\%, provides good data for about 
  22 ksec.  Due to its low energy response and the soft spectrum of the 
  pulsar's  X-ray  emission  the  LECS provides better data from PSR 
  1055$-$52 than the MECS, but doesn't contribute information on the 
  hard sources C1 and C2. 
 
\begin{figure}[h]
  \vspace{-3ex}
  \centerline{\psfig{file=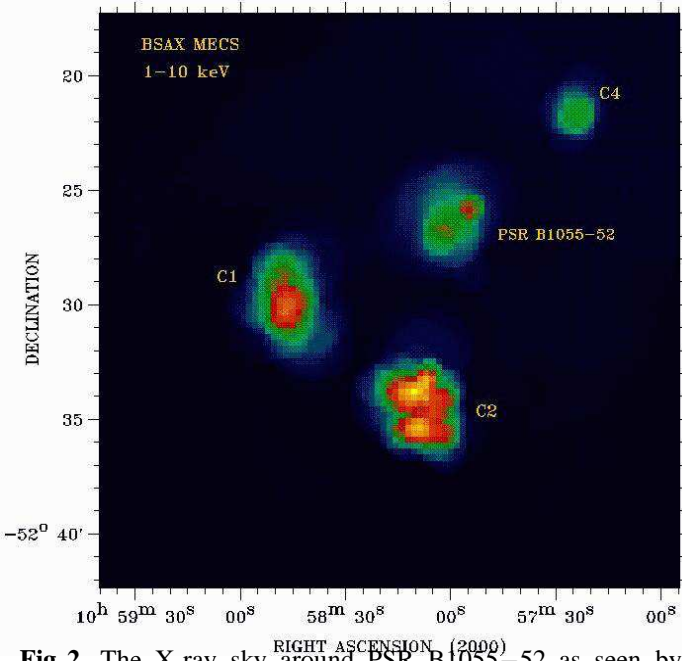,width=9.3cm,clip=}}
  \vspace{-3ex}
  \caption[h]{The X-ray sky around PSR B1055$-$52 as seen by the MECS 
   aboard BeppoSAX. Only three X-ray sources are detected in the neighborhood of
   PSR B1055$-$52. C2 is found to have a fine structure because of few close
   but spatially not fully resolved sources (see Fig.\ref{1055_pwn_rosat},
   \ref{1055_pwn_radio}). The pulsar itself is very faint beyond 1 keV.
   Its X-ray emission dominates the soft band below 0.5 keV. A new source
   C4 which is not seen in the ASCA data is detected in the north-western
   edge of the image. \label{1055_pwn_sax}\\[-4ex]}
  \end{figure}

  The  MECS image of   PSR B1055$-$52 and its   surrounding is given  in
  Fig.~\ref{1055_pwn_sax}. As shown, the  X-ray sources C1 and
  C2 neighboring  PSR 1055$-$52 are clearly  detected by the MECS, but
  appear spatially better  separated from the  pulsar than in the ASCA
  GIS data.  No diffuse and extended  component associated with C1 and
  C2  is detected. Source C2 is found   to have a fine structure,
  clearly visible in Fig.~\ref{1055_pwn_sax} and similarly seen in the
  ASCA SIS data by Shibata  et al.~(1997). In
  addition to C1, C2 and the pulsar, a new source (labeled C4), visible in
  the north-western edge   of Fig.~\ref{1055_pwn_sax}, has been detected. 

  The properties of the MECS detected  sources are summarized in Table
  \ref{sax_table}.     Since spectral fitting     of these sources  is
  precluded by the low signal-to-noise ratio we  simply have assumed a
  power-law spectrum with photon index  $\alpha=2$ for the count  rate
  to flux conversion.  The  absorption was  estimated  from the
  density of neutral Hydrogen in  the
  Galaxy (Dickey \& Lockman~1990).   Within the large  uncertainty of
  the  applied procedure, the energy fluxes  of C1 and  C2 are found to
  agree  within a factor  of two  or better with   the fluxes given by
  Shibata et al.~(1997) for the power-law model. 
  
 \begin{table*}[th!]
    \caption[]{\label{sax_table}
     Properties of the X-ray sources detected by MECS in the neighborhood of PSR B1055$-$52.\\[-3ex]}
    \begin{flushleft}
    \begin{tabular}{c c c c c c c}\hline\\[-1.5ex]
     Nr. &    Name       &  RA(2000)   &  DEC(2000)   &  
Rate$^a\!\times10^{-3}$ & Nh $\!\times10^{21}$ & flux$^b$ $\!\!\times 10^{-13}$ \\
     {}  &     BX        &     {}     &     {}      &        cts/s         & cm$^{-2}$            & erg/s/cm$^2$  \\[0.5ex]\hline\\[-1ex]
     C1  & J1058.8-5230  & 10 58 47.5  & -52 30 20    &    $3.2 \pm 0.3$     &      2.9             &  3.3 \\
     C2  & J1058.2-5233  & 10 58 10.6  & -52 33 49    &    $4.8 \pm 0.3$     &      2.9             &  4.9 \\
     C4  & J1057.8-5221  & 10 57 25.2  & -52 21 33    &    $2.5 \pm 0.2$     &      2.9             &  2.6 \\[1ex]\hline\\[-4ex]
 \end{tabular}
  \end{flushleft}
   $^a$background, dead-time and vignetting corrected, $^b$in the $0.5-10$ keV band pass
 \end{table*}

  \begin{figure}[h]
  \begin{picture}(92,84)
  \put(0,1){\psfig{file=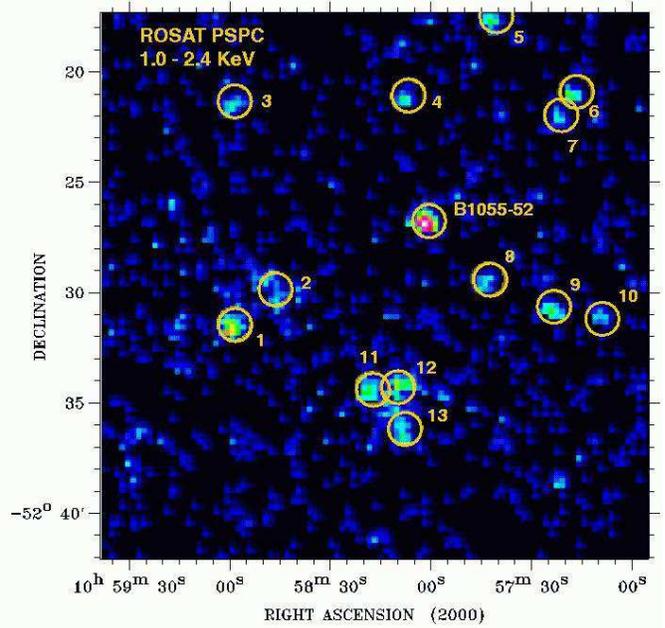,width=8.8cm,clip=}}
  \end{picture}
  \caption[h]{PSR B1055$-$52 as seen by the ROSAT PSPC. The image has the
   same scale as the BeppoSAX MECS image shown in Fig.\ref{1055_pwn_sax}.
   Several faint X-ray sources are detected in the pulsar's neighborhood.
   The position of source \#2 and \#11$-$13 is in agreement with the location
   of the X-ray sources C1 and C2 observed by ASCA and MECS (i.e.~C2 is
   resolved into 3 point sources by ROSAT). Source \#3 corresponds to
   "clump C3" of Shibata et al.~(1997).\label{1055_pwn_rosat}}
  \end{figure}

  Whereas the instruments  abord ASCA and BeppoSAX provide information
  over  a larger  energy  range, the ROSAT   instruments  have the better
  angular  resolution  and soft  response.  PSR B1055$-$52 was observed
  with the ROSAT  PSPC in January 1992 for  an effective exposure time
  of 15  ksec. Also  data from  a  9 ksec  ROSAT HRI   observation are
  available in the ROSAT  archive, but the analysis  of these data did
  not add new information  to  the PSPC results. Furthermore,  the
  PSPC's background  is much lower than  that of the HRI  detector, so
  that the PSPC is   the preferable  instrument  to search  for  faint
  and/or diffuse emission. 

  Using  the    Extended    Scientific   Analysis     Software    {\em 
  e\raisebox{-.22ex}{X}sas} (Zimmermann et al.~1994) we have performed a
  spatial analysis of the  pulsar field in  the soft (0.1$-$1.0 keV) and
  in the hard band (1.0$-$2.4 keV).  13 sources are detected by the PSPC
  with a significance $\ge  5 \sigma$  at off-axis angles  $\le  20$ arcmin, 
  as this field includes the  full GIS field  of  view.  The properties of
  these 13  sources  are summarized in Table~\ref{1055_rosat_table}.
  The PSPC hard  band   image with  the sources  encircled  is  shown in
  Fig.~\ref{1055_pwn_rosat}.

\begin{table*}[ht!]
    \caption[]{Properties of the ROSAT PSPC sources, detected in the neighborhood of PSR B1055$-$52 beyond 1 keV
     and within an off-axis angle of $\le 20$ arcmin. The given energy range indicates if a source is detected
     in the full band or beyond 1 keV only. The columns "mag" and "ID" give the estimated R/B-band magnitude
     and the proposed identification of the optical counterparts found in a 10 arcsec error circle centered
     on the PSPC position. "S" and "nS" indicate if the sources in the error circle are likely to be
     identified with a star or with not being a star.\label{1055_rosat_table}\\[-6ex]}
    \begin{flushleft}
    \begin{tabular}{c c c c c c c c c c c c}\hline\\[-1.5ex]
     Nr. &  Range  & Name          &  ASCA  & RA(2000)  &  DEC(2000)  &   \Rate        &   \Nh      &    \flux      &   mag  & mag   & ID  \\
     {}  &   keV   &  RX           &   ID   &   {}      &    {}       &    cts/s       &  cm$^{-2}$ &  erg/s/cm$^2$ &    R   &  B    & {}  \\[0.4ex]\hline\\[-1ex]
     1   & 0.1-2.4 & J1058.9-5231  &   {}  & 10 58 58.1 & -52 31 36.8 &  $4.6 \pm 0.6$ &     2.9    &     2.3       &   &   &   \\
     2   & 1.0-2.4 & J1058.7-5229  &   C1  & 10 58 46.9 & -52 29 48.5 &  $2.7 \pm 0.4$ &     2.9    &     2.1       & $>22$   &$>23$   & \\
     3   & 1.0-2.4 & J1058.9-5221  &   C3  & 10 58 57.5 & -52 21 23.1 &  $0.8 \pm 0.3$ &     2.9    &     0.6       &  $>$22   &  $>$23   & {}  \\
     4   & 0.1-2.4 & J1058.0-5221  &   {}  & 10 58 05.6 & -52 21 08.6 &  $1.6 \pm 0.4$ &     2.9    &     1.3       &   {}   &  {}   & {}  \\
     5   & 0.1-2.4 & J1057.6-5217  &   {}  & 10 57 39.4 & -52 17 26.8 &  $2.7 \pm 0.5$ &     3.3    &     1.5       &   {}   &  {}   & {}  \\
     6   & 0.1-2.4 & J1057.2-5220  &   {}  & 10 57 15.9 & -52 20 55.4 &  $2.1 \pm 0.5$ &     3.3    &     1.1       &   {}   &  {}   & {}  \\
     7   & 0.1-2.4 & J1057.3-5222  &   {}  & 10 57 20.6 & -52 22 03.7 &  $1.7 \pm 0.5$ &     3.3    &     1.6       &   {}   &  {}   & {}  \\
     8   & 1.0-2.4 & J1057.6-5229  &   {}  & 10 57 41.8 & -52 29 31.7 &  $0.8 \pm 0.3$ &     3.3    &     0.7       &   {}   &  {}   & {}  \\
     9   & 0.1-2.4 & J1057.3-5230  &   {}  & 10 57 22.1 & -52 30 48.6 &  $1.8 \pm 0.5$ &     3.3    &     1.0       &   {}   &  {}   & {}  \\
    10   & 0.1-2.4 & J1057.1-5230  &   {}  & 10 57 06.8 & -52 30 59.3 &  $2.4 \pm 0.5$ &     3.3    &     1.3       &   {}   &  {}   & {}  \\
    11   & 0.1-2.4 & J1058.2-5234  &   {}  & 10 58 16.4 & -52 34 28.1 &  $2.0 \pm 0.5$ &     2.9    &     1.0       &  13.7  & 13.9  & nS  \\
    {}   &   {}    &     {}        &   {}  &  {}      &     {}      &        {}      &     {}     &     {}        &  21.5  &  22.2   & nS  \\[-5ex]
    12   & 0.1-2.4 & J1058.1-5234  & $\!\!\!\!\!\!\!\!\left\}\begin{array} {c c c c c c} {} \\ {} \\ {} \\{} \\{} \end{array} \right.\!$
                                       C2  & 10 58 08.0 & -52 34 27.1 &  $4.8 \pm 0.8$ &     2.9    &     \quad\quad$\;\!$2.4$\;\;\;\;\left\{
       \begin{array} {c c c} {} \\ {} \\{} \end{array} \right.\!\!\!\!\!$ &  18.2  & 18.0  & nS  \\[-5.0ex]
              {}   &   {}    &     {}        &   {}  & {}      &     {}      &        {}  &     {}     &     {}        &  14.3  & 14.1  & nS  \\
    13   & 0.1-2.4 & J1058.1-5236  &   {}  & 10 58 06.6 & -52 36 24.1 &  $3.3 \pm 0.6$ &     2.9    &     1.6       &  18.4  & 16.2  & nS  \\[1ex]\hline\\[-4ex]
 \end{tabular}
 \end{flushleft}
 $^a$background, dead-time and vignetting corrected rate in the given range,\newline
 $^b$within the $0.5-10$ keV band pass,\newline
\end{table*}                                

  Correlating the position of the sources  detected by ROSAT, ASCA and
  SAX, we have identified the PSPC source \#2 as the likely counterpart 
  of the hard X-ray source C1 and the sources \#11$-$13 as the counterparts 
  of C2.  Source \#3 is found to correspond to the "clump" C3  of
  Shibata et  al.~(1997) which is not  detected  in the  MECS
  data. The PSPC sources  \#6,7 correspond to  source C4 seen  in the
  MECS  image (see Fig.~\ref{1055_pwn_sax}).  The energy fluxes of the
  PSPC sources are  in agreement with what  we have measured for their
  MECS  counterparts by   adding  the  fluxes  from  source  \#6,7 and
  \#11-13, respectively. 
  
\begin{figure}[h!]
 \vspace{2ex}
 \centerline{\psfig{file=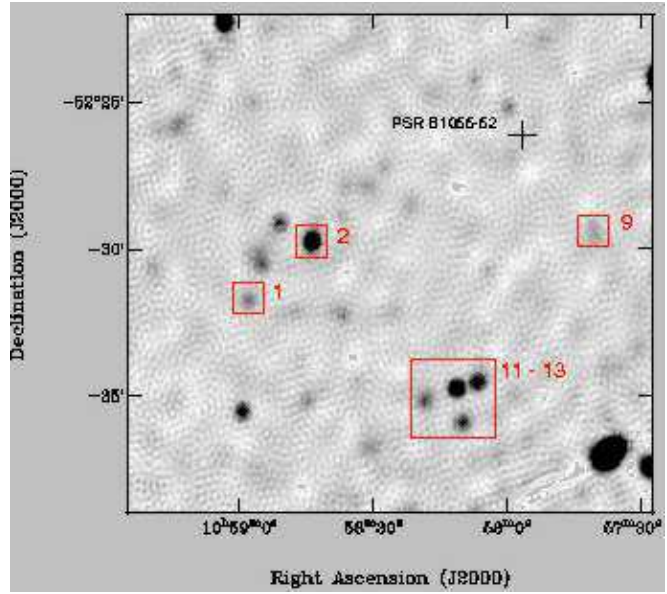,width=8.7cm,clip=}}
  \caption[h]{1.4 GHz radio image of PSR 1005$-$52 and its surrounding. The
  image has been adapted from \cite{Stapper_et_al99}(1999). Radio counterparts of
  the ROSAT detected sources \#1,2,9 and \# 11-13 are indicated. The sources
  \#2 and \#11$-$13 correspond to the ASCA and MECS detected hard X-ray sources
  C1 and C2. \label{1055_pwn_radio}}
  \end{figure}

  Our analysis has thus shown, that the hard X-ray sources C1 and C2, which
  are proposed to  be caused by the interaction of a  pulsar-wind with
  the interstellar medium, are resolved by  the PSPC into single point 
  sources, well  separated from  the  pulsar and unlikely to be
  associated with it. 
  A  result,  which is in  line with  a  report  recently published by
  \cite{Stapper_et_al99}(1999).  The  authors    have performed    radio
  observations  of the field around PSR  B1055$-$52 in order to search
  for diffuse  and  extended radio counterparts  of  the putative ASCA
  pulsar-wind  nebulae.  Their  observations,  however, have failed to
  detect extended plerionic emission associated with PSR B1055$-$52 (on
  any spatial scale),  but have shown  that the  position of  the ASCA
  knots  C1  and    C2   correlate with  (unidentified)   radio  point
  sources. The radio image of the pulsar  field, adopted from 
  \cite{Stapper_et_al99}(1999) but with  their  contour lines  removed, 
  is  shown in Fig.\ref{1055_pwn_radio}.  The  radio   counterparts which    
  we have identified to correspond to the sources C1, C2 and the PSPC 
  sources \#1 and \#9 are indicated by squares.

  A search for possible optical  counterparts within a 10 arcsec error
  circle of the ROSAT sources \#2,3 and \#11$-$13 was performed making
  use of results obtained   from the pipeline processing of digitized
  Schmidt photographic plates, used as input for the generation of the
  Guide Star Catalog II (McLean et al.~1998).  The plates are scanned
  at $15  \mu$ resolution, corresponding  to a pixel  size of 1 arcsec
  and  the astrometric  calibration  (accurate  to $\sim 0.5$  arcsec,
  absolute) has  been performed using Tycho/Hipparcos reference stars.
  Conversion from photographic density to B and R magnitudes has been
  achieved  using inputs  from the  Guide  Star Photometric  Catalogue2
  (Postman et al.~1998) and are accurate within 0.2 magnitudes. 
  Star/non-star classification is based on the object morphology as
  characterized by the pipeline. The result of our correlation is included in
  Table~\ref{1055_rosat_table}.  We  have found a candidate   optical
  counterpart for  each of the sources  \#2, \#11, \#13 while for
  \#12  we  have found three objects within    the   error circle.
  Although from inspecting the shape of the optical sources we
  find that they are unlikely to be  stars, their final identification
  is subject of subsequent observations.

\subsection{PSR B0656+16 \label{0656}}

  The 384 ms pulsar PSR  B0656+14 turns out  to be very similar to PSR
  B1055$-$52  (see Table 3 of  Becker \& Tr\"umper  1997). Its inferred
  pulsar  parameters  are a spin-down  age of  $1.1\times  10^5$ yr, a
  magnetic  dipole  field  $B_\perp  =   4.7\times  10^{12}$ G  and  a
  spin-down   energy  $\log \dot{E}    =   34.58$ erg/s.   The pulsar
  dispersion measure, based on the Galactic  electron density model of
  Taylor \& Cordes (1993), puts PSR B0656+14  at a nominal distance of
  760 pc. 

  X-ray  emission  from PSR B0656+14  was first  seen by  the {\em
  Einstein} observatory    (Cordova  et  al.~1989).   ROSAT Observations
  detected X-ray pulses  at the radio period  and identified  the  soft
  X-ray  emission  as radiation  from  the hot cooling  neutron  star
  surface (Finley et al.~1992; Possenti et al.~1997). 
  In addition to the soft  X-ray emission, a hard  tail component in the
  X-ray spectrum, indicated from  the ROSAT data, was  observed by ASCA and
  identified to be of non-thermal origin (Greiveldinger et al.~1996). 

  The possible existence  in the ASCA data of  a $\sim 20$ arcmin wide
  putative   pulsar-wind   nebula   around  PSR   B0656+14   has  been
  communicated  by  Kawai \& Tamura (1996) and by Kawai et al.~(1998a, 1998b) 
  but no detailed information about its flux and size is available in the 
  literature. 
  In order to get the best inputs for  the search of possible ROSAT
  counterparts of the hypothetical  pulsar wind nebula, we re-analyzed 
  the archival ASCA data of PSR B0656+14. 
  Since  the   data from  the  SIS   were not found   to   add further
  information, we restricted our analysis to the GIS2 and GIS3 data. 
  To optimize the detectors temporal resolution for pulsar timing, the
  GIS2   and GIS3 were operated  in  a special   PH  mode with reduced
  spatial resolution (raw pixel size 1 arcmin only). 
  Applying the standard screening  criteria (no rise-time  information
  was   recorded during the   observations  so  that  a cleaning   for
  cosmic-ray/particle events could  not be performed) we obtained good
  data from GIS2  and GIS3 for  an effective exposure  time of 35\,807
  and 35\,806 sec, respectively. 
  Important to note is that in this observation the detector aim-point
  was  placed at the so-called  "1CCD nominal position", which is near
  the  center of S0  chip-1 and  S1 chip-3 of  the SIS  detector. This
  configuration implies  for the GIS data  that the pulsar is observed
  near the optical axis in GIS3 but approximately 5 arcmin off-axis in
  GIS2 (vice versa for parts of the pulsar surrounding). 
   
  Since this configuration results in  a different sensitivity map for
  GIS2  and GIS3\footnote{Both instruments  are  most sensitive at the
  optical  axis and noticeable less  sensitive 5 arcmin away from it.}
  and both detectors have a different energy response, we have decided
  to analyse the images from GIS2 and GIS3 separately.
 
  \begin{figure}[t] 
   \begin{picture}(88,160)(0,-2)
   \put(0,80){\psfig{file=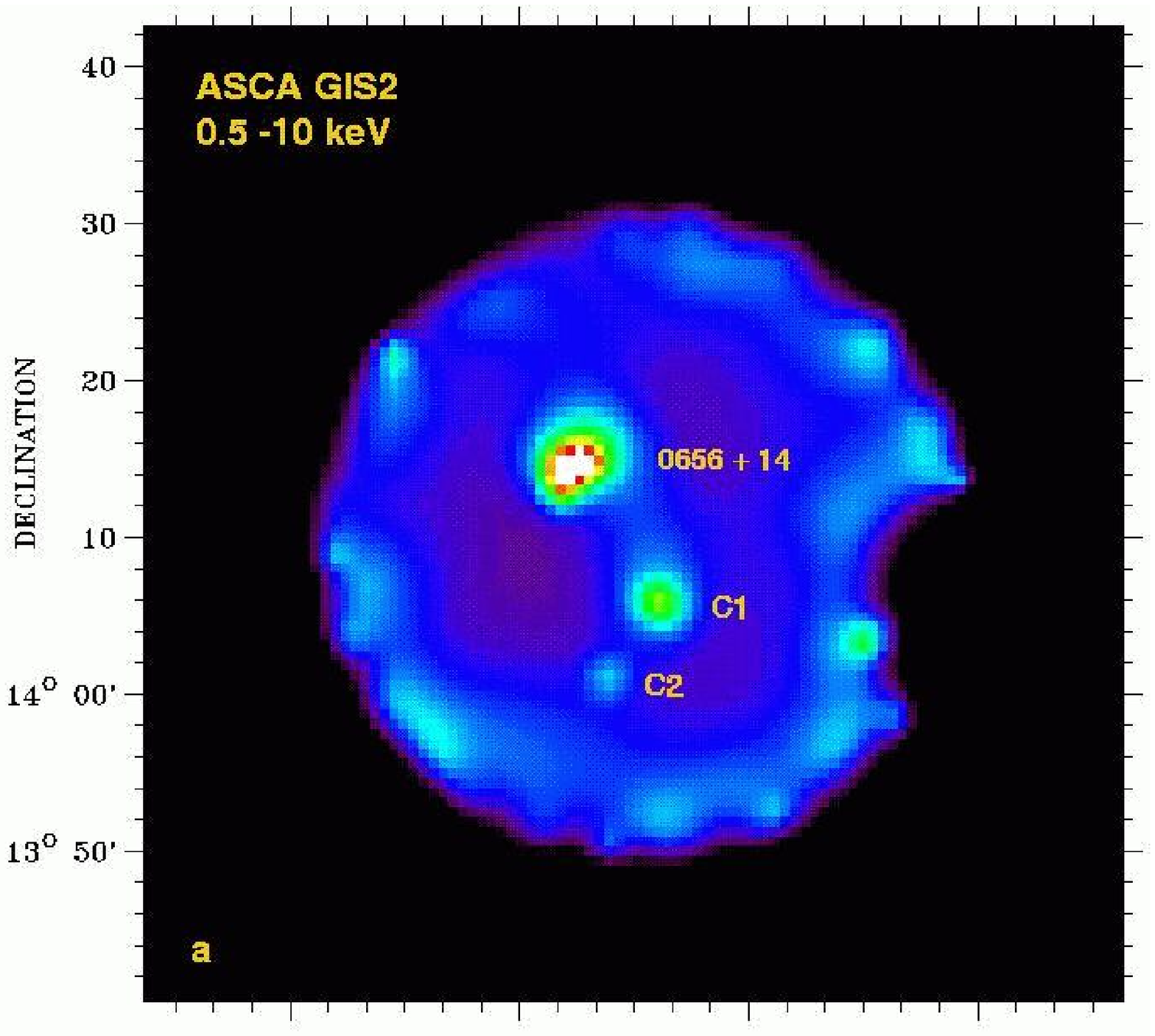,width=8.8cm,clip=}}
   \put(0,-7){\psfig{file=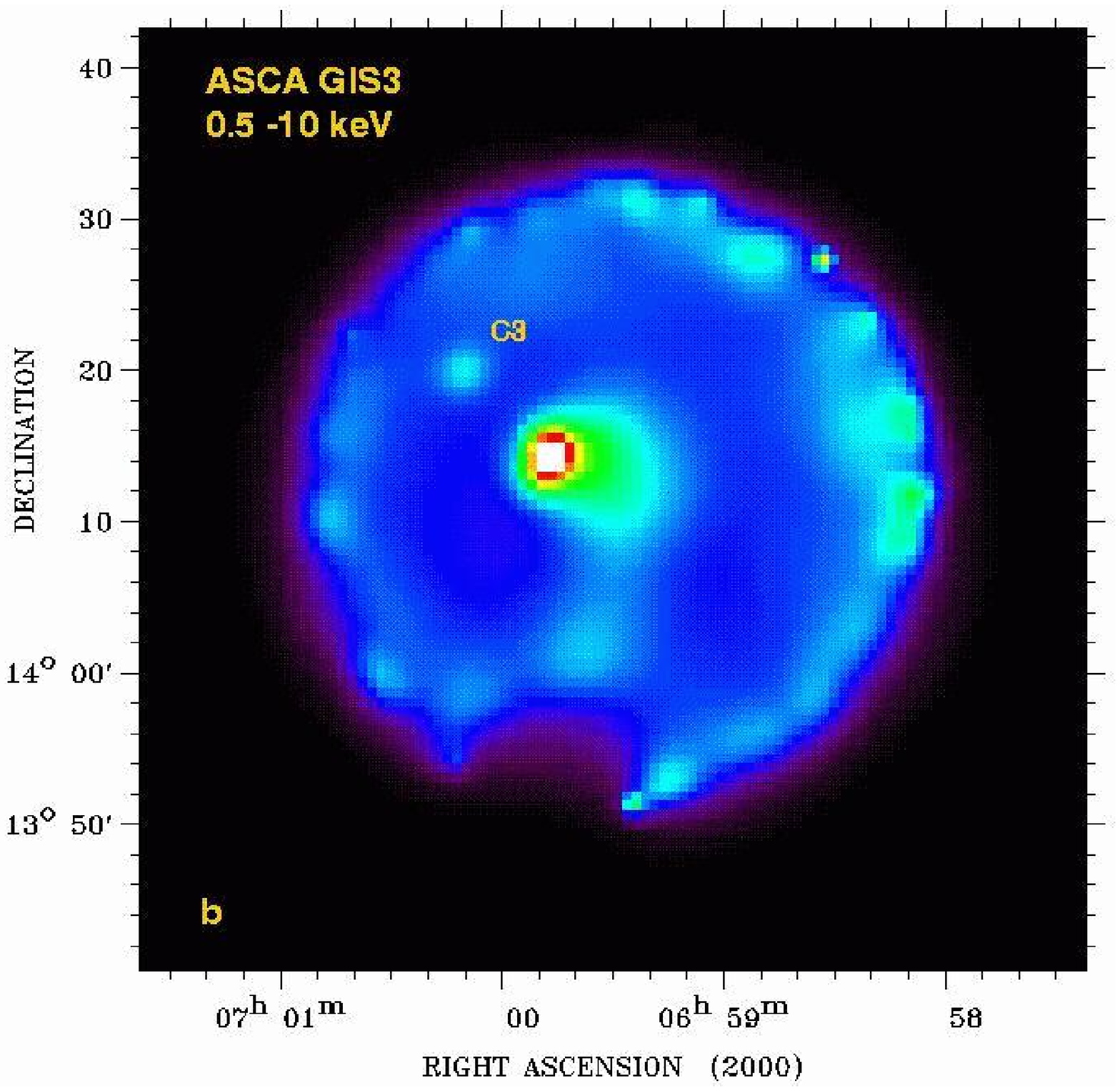,width=8.8cm,clip=}}
   \end{picture} 
   \vspace{2ex} 
   \caption{{\bf a,b}\quad PSR B0656+14 and
   its   surrounding  as  seen  with  the ASCA    GIS2 (top)  and GIS3
   (bottom). Shown is the GIS central $\sim 44$ arcmin field of view
   (images are not vignetting  corrected). The pulsar is the brightest
   source in the  field. Two faint but hard  sources C1 and C2 show up
   close  to  the  GIS2  optical  axis.   These sources are  observed
   off-axis by  the GIS3 and are barely  visible. A  third faint point
   source   C3   shows up  in   the north-western  direction  from the
   pulsar. In both images there is no  indication  for a  pulsar-wind 
   nebula. The  apparently brighter emission  at  the  edges of  the
   detector fields appears because most non X-ray background events in
   the  GIS  are registered  close  to    the   detector     walls.}
   \label{0656_pwn_asca_gis2}\label{0656_pwn_asca_gis3} 
   \end{figure} 
            
  In order to visually enhance  faint (and extended) structures in the
  noise-dominated GIS data we have processed the extracted images with
  an adaptive kernel smoothing (Ebeling et al.~1999). The advantage of
  this  kind  of   image  processing compared  to   a  simple Gaussian
  smoothing    (which    can    produce   strong    artifacts   easily
  miss-interpreted as pulsar-wind nebula)  is that noise is suppressed
  very efficiently while at the same time  real structure is preserved
  on all scales. 

  Fig.~\ref{0656_pwn_asca_gis2}a   shows the  GIS2 image,  with the
  pulsar  observed off-axis.  Two faint  point  sources C1 and  C2 are
  clearly  visible south  of the  pulsar.  The  brighter  source C1 is
  closest to the detector optical  axis where the instrument is most
  sensitive. There is  no indication for  the existence of a $\sim 20$
  arcmin wide  clumpy pulsar-wind nebula   in this  data. Indeed, the
  putative  pulsar-wind nebula, which Kawai \& Tamura (1996) found to
  have a shape like the  "trunk of an  elephant", actually consists of
  two  point sources.  This  is confirmed by   the GIS3 image shown in
  Fig.\ref{0656_pwn_asca_gis3}b, where  the pulsar is observed on-axis
  and the two faint sources C1 and C2 are about 5 arcmin off-axis.  At
  this  position,  the  sensitivity and   spatial  resolution  of  the
  detector are noticeably lower than  on-axis, so that the sources  C1
  and  C2 are only barely   visible/resolved.  A third source (C3)  is
  also visible in the GIS3 image, north-west of  the pulsar. Due to
  their low signal-to-noise ratios, no spectral analysis is possible on
  the sources C1$-$C3. 

\begin{figure}[t]
  \begin{picture}(88,240)(0,0)
  \put(5,161){\psfig{file=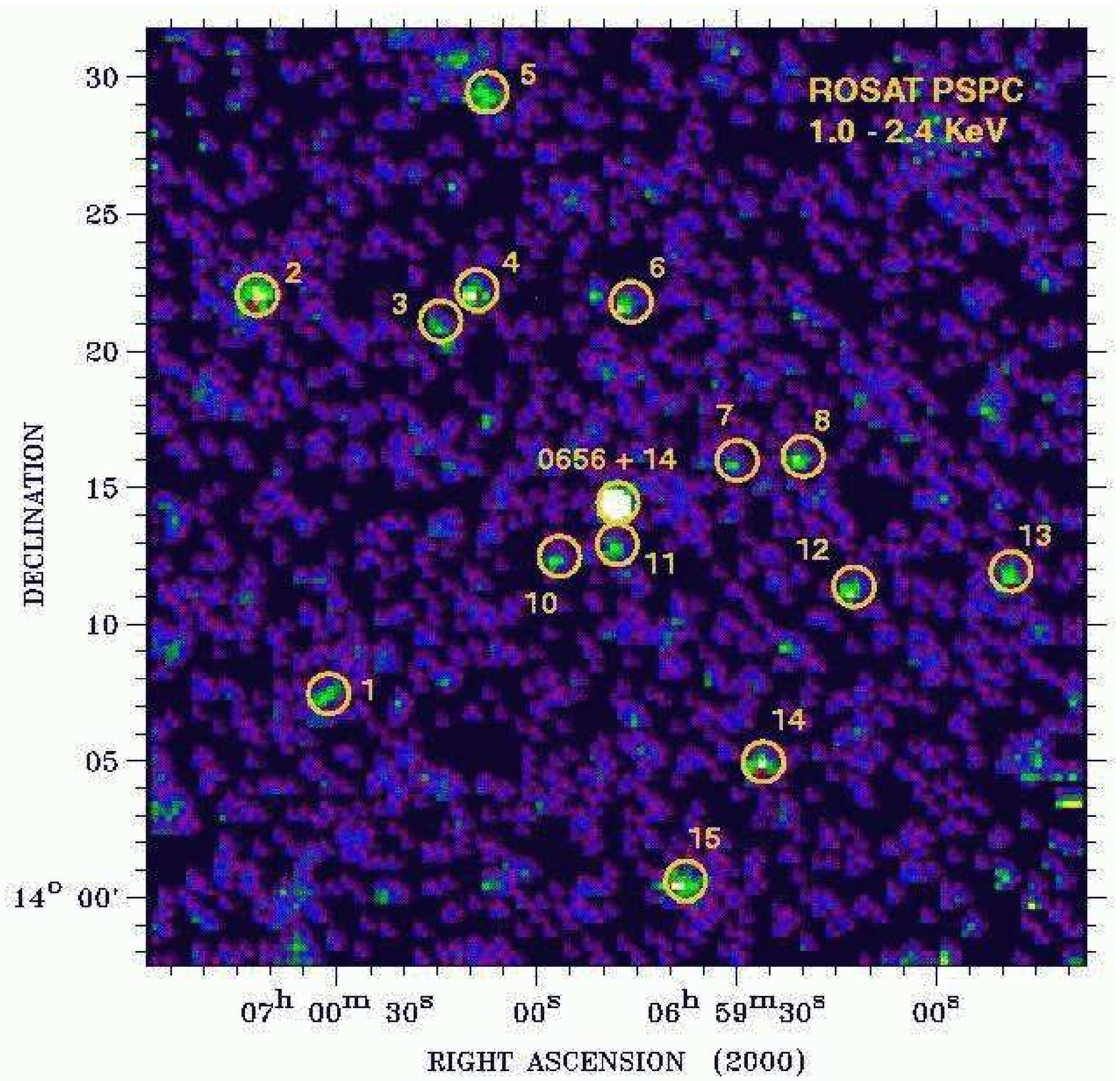,width=8cm,clip=}}
  \put(0,0){{}}
  \end{picture}
  \vspace{-16.2cm}
  \caption[h]{The PSR B0656+14 field as seen by the ROSAT PSPC. The pulsar is the
   bright central point source. The position of  source \#14,15 and \#3
   is in agreement with the location of the X-ray sources C1$-$C3 observed
   by the ASCA GIS. The position of source \#11 corresponds to a slightly extended
   radio source discussed in Cordova et al.~(1989).}
  \label{0656_pwn_rosat}
  \end{figure}

  Since BeppoSAX observed PSR  B0656+14 only in  1999 January and  the
  data are not yet  available to us, we  concentrate our search for soft
  X-ray  counterparts of C1$-$C3  using the  ROSAT data only. PSR
  B0656+14 was observed by ROSAT with both the PSPC and HRI detectors in
  March 1991 and 1992 for an effective  exposure time of  16 ksec and 10
  ksec, respectively. 
  Spatial analysis of  the PSPC field, restricted  to  an off-axis angle
  $\le 20$   arcmin, detected  15 sources,   at a  significance $\ge  5
  \sigma$. 
  Apart from the pulsar itself, all  sources  are rather faint with
  typical  counting   rates   of   $(3-8) \times   10^{-3}$   cts/s (see
  Table~\ref{0656_rosat_table} for a summary of their properties).  A
  spectral analysis is therefore precluded by the low number of detected
  counts. The PSPC hard band image,  showing the central inner region of
  the PSPC  detector with the  detected  sources encircled, is  shown in
  Fig.~\ref{0656_pwn_rosat}. 

  \begin{table}[b]
    \parbox{18cm}
    {\caption[]{Properties of the  ROSAT PSPC sources, detected beyond 1
     keV within an region of $\le  20$ arcmin around PSR B0656+14. The
     given energy range indicates if a source is  detected in the full
     band or beyond  1 keV only. The columns  "mag" and  "ID" give the
     estimated B-band magnitude and the proposed identification of the
     optical  counterparts found in a 10  arcsec error circle centered
     on the PSPC position. "nS" indicate that the  source in the error
     circle is likely to be identified with not being a star. \label{0656_rosat_table}}}\\[-6ex] 

    \begin{flushleft}
    \begin{tabular}{c c c c c c c c c c}\hline\\[-1.5ex]
     Nr. &  Range  & Name          & ASCA & RA(2000)   &  DEC(2000)   & \Rate       &  mag   & mag & ID   \\
     {}  &   keV   &  RX           &   ID &    {}     &      {}       &  cts/s      &  B  & R  & {}   \\[0.5ex]\hline\\[-1ex]
     1   & 0.1-2.4 & J0700.5+1407  &   {} & 07 00 32.6  & +14 07 17.6 & $3.22 \pm 0.60$ &   {} &   {}  & {}   \\
     2   & 0.1-2.4 & J0700.7+1422  &   {} & 07 00 42.8  & +14 22 03.7 & $7.02 \pm 0.87$ &   {} &   {}  & {}    \\
     3   & 1.0-2.4 & J0700.2+1421  &   C3 & 07 00 15.0  & +14 21 03.7 & $5.93 \pm 0.21$ & 21.6 & 22.3  & nS   \\
     4   & 0.1-2.4 & J0700.1+1422  &   {} & 07 00 09.5  & +14 22 11.9 & $4.02 \pm 0.62$ &   {} &   {}  & {}   \\
     5   & 0.1-2.4 & J0700.1+1429  &   {} & 07 00 08.1  & +14 29 27.9 & $6.34 \pm 0.83$ &   {} &   {}  & {}   \\
     6   & 0.1-2.4 & J0659.7+1421  &   {} & 06 59 45.8  & +14 21 54.7 & $2.43 \pm 0.52$ &   {} &   {}  & {}   \\
     7   & 1.0-2.4 & J0659.4+1415  &   {} & 06 59 29.7  & +14 15 57.5 & $0.59 \pm 0.21$ &   {} &   {}  & {}   \\
     8   & 0.1-2.4 & J0659.3+1416  &   {} & 06 59 19.6  & +14 16 06.1 & $2.26 \pm 0.49$ &   {} &   {}  & {}   \\
     9   & 0.1-2.4 & PSR B0656+14  &   {} & 06 59 48.1  & +14 14 25.4 & $1897.24\pm 10.9$ & {} &   {}  & {}   \\
    10   & 0.1-2.4 & J0659.9+1412  &   {} & 06 59 56.5  & +14 12 25.7 & $6.32 \pm 0.80$ &   {} &   {}  & {}   \\
    11   & 1.0-2.4 & J0659.8+1412  &   {} & 06 59 48.0  & +14 12 50.0 & $0.83 \pm 0.25$ &   {} &   {}  & {}   \\
    12   & 0.1-2.4 & J0659.1+1411  &   {} & 06 59 11.8  & +14 11 18.1 & $4.46 \pm 0.66$ &   {} &   {}  & {}   \\
    13   & 0.1-2.4 & J0658.7+1411  &   {} & 06 58 47.8  & +14 11 52.4 & $2.93 \pm 0.63$ &   {} &   {}  & {}   \\
    14   & 0.1-2.4 & J0659.4+1404  &   C1 & 06 59 25.4  & +14 04 44.2 & $3.70 \pm 0.62$ & 19.4 & 20.8  & nS   \\
    15   & 0.1-2.4 & J0659.6+1400  &   C2 & 06 59 37.7  & +14 00 18.9 & $8.18 \pm 0.99$ & 15.7 & 16.1  & nS   \\[1ex]\hline\\[-4ex]
  \end{tabular}
  \end{flushleft}
  \parbox{18cm}{$^a$background, dead-time and vignetting corrected rate in the given energy range.}
 \end{table}

  After a spatial correlation between the PSPC and the ASCA image 
  we identified   the ROSAT sources    \#14,15 and \#3   as the likely
  counterparts   of the  clumps  C1$-$C3.   In   order to improve the
  positions of the PSPC sources (accurate within $\simeq 10$ arcsec,
  including systematic   errors like   pointing inaccuracies   of  the
  satellite), we have analyzed archival  HRI data. However, due to the
  3-5 times lower sensitivity of the HRI with respect to  the PSPC and 
  the shorter exposure, we found that only  source \#15 is marginally 
  detected, with a position (RA=06:59:38.26 and  DEC= +14:00:30.8) in  
  agreement with the PSPC one (see Table \ref{0656_rosat_table}). Again, 
  a search for optical counterparts within  10 arc-

\newpage
\noindent
  sec error  circles of the  ROSAT sources \#14,15 and \#3 (i.e.~C1$-$C3) 
  was performed using digitized POSS-II B  and R plates. The result of our 
  correlation is  also included in  Table \ref{0656_rosat_table}. For each 
  of the interesting sources,  we  have found candidate optical counterparts. 
  Although the nature of the ROSAT sources \#14,15 and \#3 has still to 
  be assessed, our analysis suggests that they are the counterparts 
  of the emission clumps observed by ASCA and that they are background 
  objects unrelated to the pulsar.

\begin{figure}[h]
  \begin{picture}(80,80)(0,0)
  \put(0,0){\psfig{file=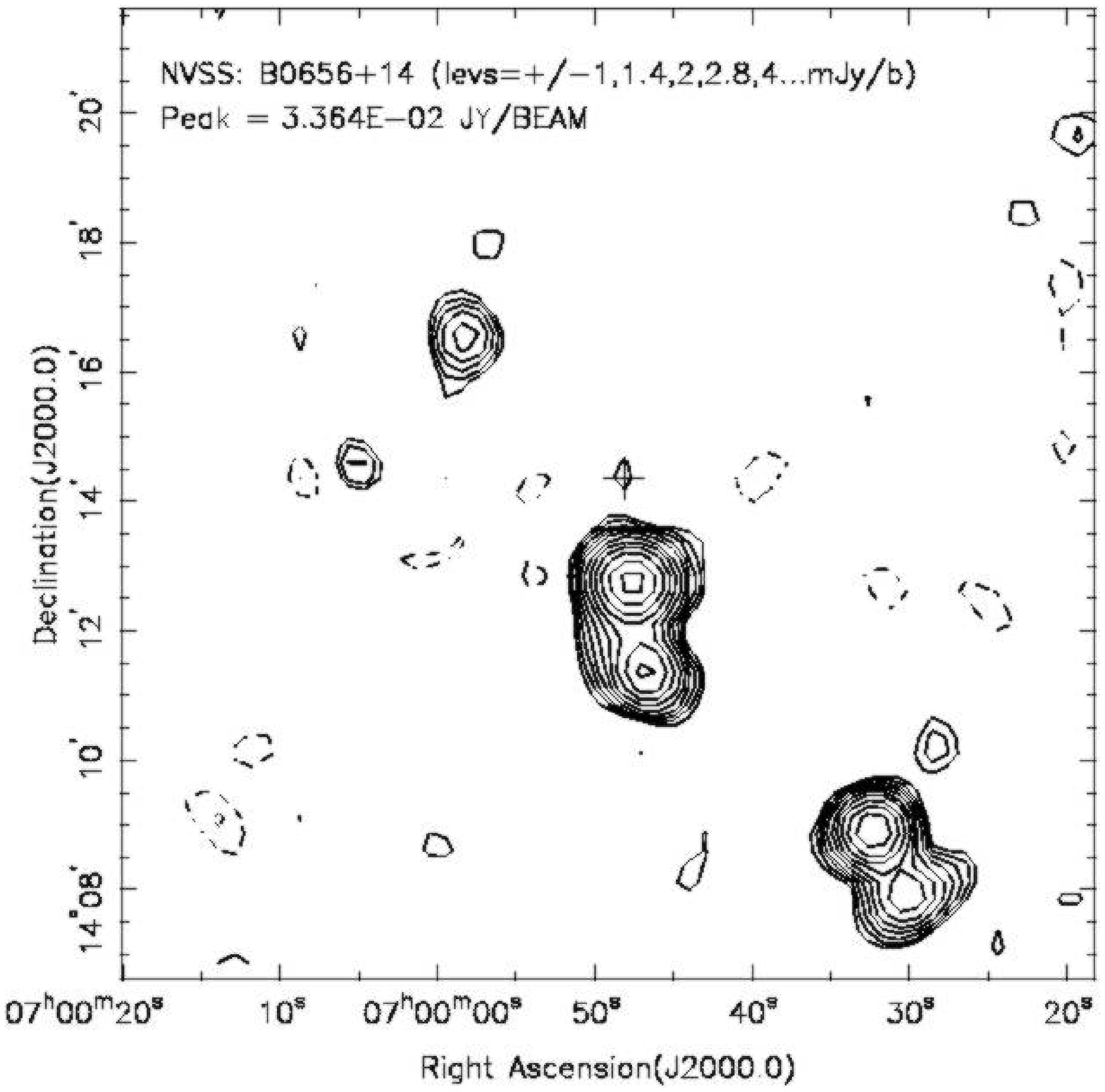,width=8cm,clip=}}
  \end{picture}
  \caption[h]{The PSR B0656+14 field as seen in the NVSS (Condon et al.~1998). 
   The location of the pulsar is marked by a cross. The close by radio source   
   $\sim 1.5$ arcmin south of the pulsar is observed to be 40-50\% polarized 
   (Cordova et al.~1989) and shows up in our correlation to be the radio 
   counterpart of the PSPC source \#11.\label{0656_radio}}
  \end{figure}

  In the radio band a search for a pulsar-wind nebula surrounding PSR B0656+14
  was originally performed by Cordova et al.~(1989) using the VLA. The
  authors  reported on a candidate source about $1.2-3.3$ arcmin away
  from the pulsar (Fig.~\ref{0656_radio}), which appeared slightly 
  extended and about $40-50\%$ linear polarized. This was taken as a 
  clear sign that the radio  emission of this  source   is due to  
  synchrotron radiation (as  expected for a  plerion). Our analysis 
  shows that the extended  radio source of Cordova et al.~(1989) 
  coincides with the position of the hard PSPC  source  \#11, although 
  undetected by the Einstein Observatory.  Whether  this radio/X-ray  
  source is really related to the pulsar or whether it is an unrelated 
  background object remains to be subject to further studies. 

\subsection{Geminga}

  The Geminga pulsar  was first noticed  as  a strong gamma-ray  source
  (\cite{Fichtel_et_al75}1975)  and  only recently    found to   be  a
  rotation-powered         pulsar         (\cite{Halpern_Holt_92}1992;
  \cite{Becker_et_al93}1993). It  has  a  spin-down age  of $3.4\times
  10^5$     yr,    an    inferred   magnetic   dipole    component  of
  $B_\perp=1.6\times  10^{12}$ G and    a  spin-down energy of   $\log
  \dot{E}=34.51$   erg/s     (Bertsch  et   al.~1992). Its distance of
  $159^{+59}_{-34}$   pc   was   determined  from    optical  parallax
  measurements using the HST (Caraveo et al.~1996). 

  X-ray emission from Geminga was first detected by the {\em Einstein}
  observatory (Bignami et al.~1983). However,  as for PSR B1055$-$52, the
  sensitivity of ROSAT was required to detect its  X-ray pulses and to
  identify its soft  X-ray emission as radiation from the hot cooling
  neutron star surface (Halpern  \& Holt  1992).  A harder  spectral
  component,  first  interpreted   as thermal   emission  from  a  hot
  polar-cap, was later identified   with  ASCA to be   of  non-thermal
  (magnetospheric)  origin (Halpern \& Wang  1997;  see also Becker \&
  Tr\"umper 1997 and the discussion therein). 
  The possible existence
  of a  $\sim 20$ arcmin  wide  pulsar-wind nebula  around Geminga was
  first reported by Kawai et al.~(1998a, 1998b). 

\begin{figure*}[th]
   \begin{picture}(160,90)(0,0)
   \put(0,0){\psfig{file=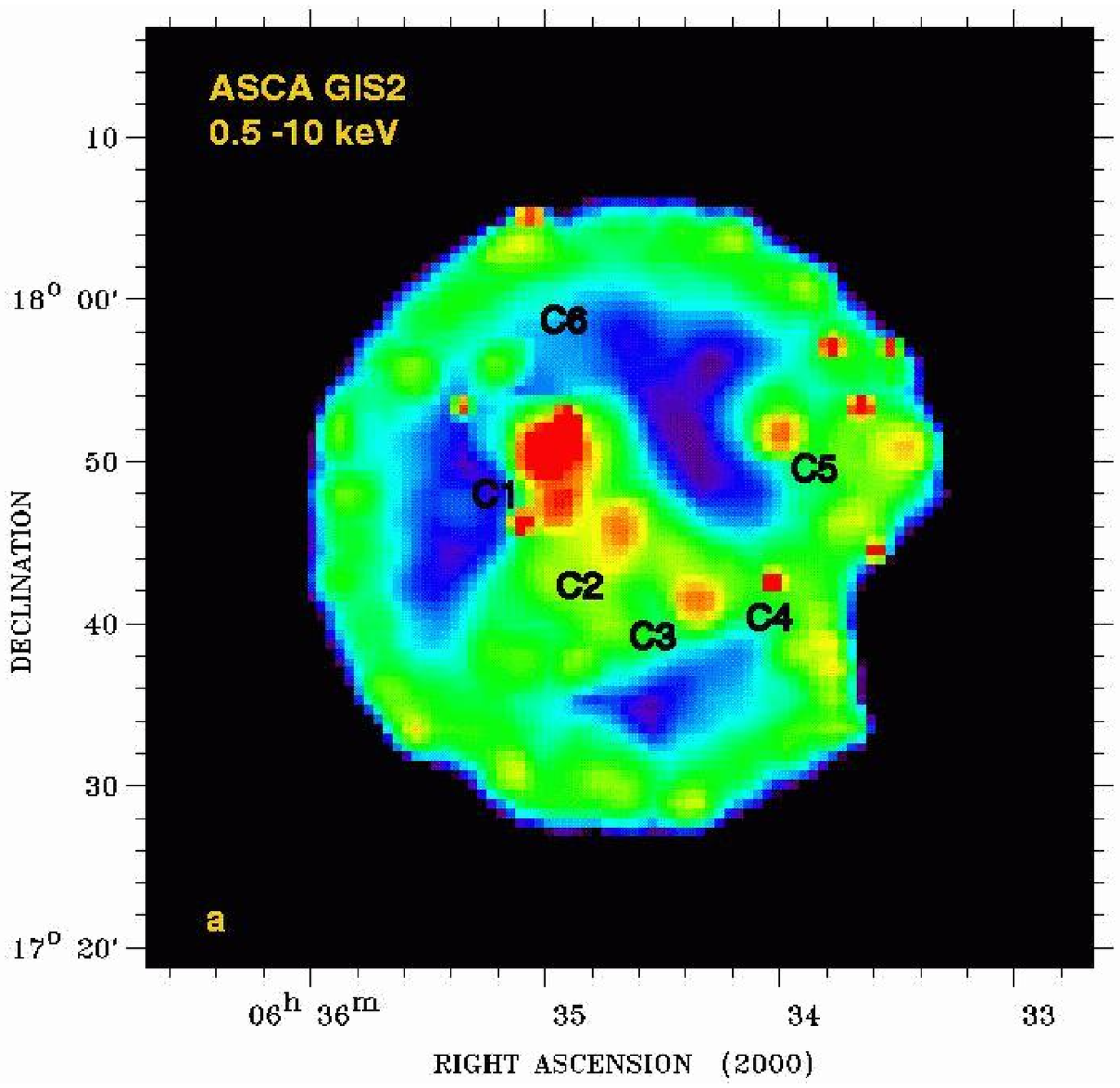,width=8.8cm,clip=}}
   \put(92,0){\psfig{file=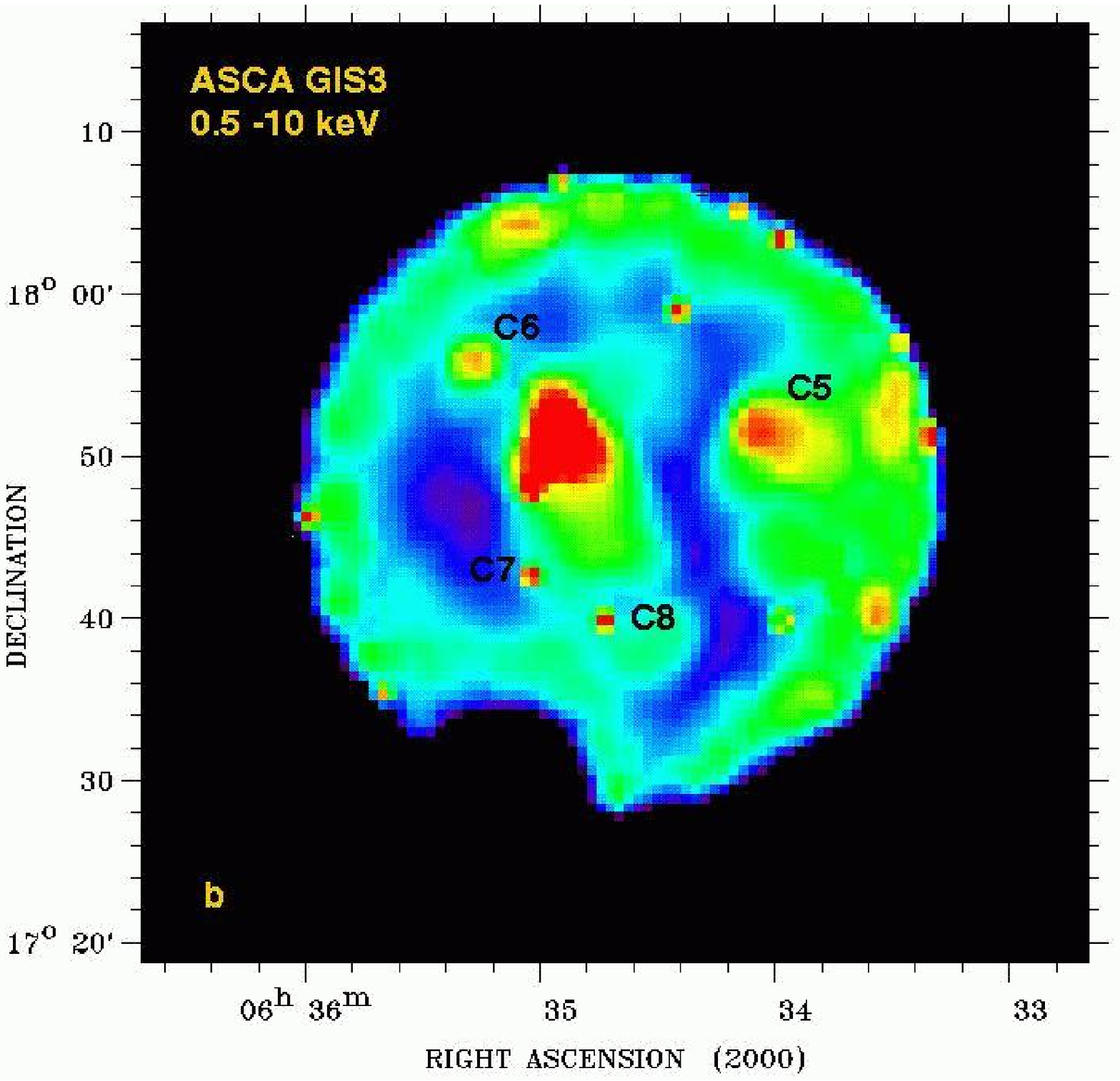,width=8.8cm,clip=}}
   \end{picture}
   \caption{{\bf  a,b}\quad Geminga  and its surrounding  as seen with
   the ASCA  GIS2 (top) and GIS3  (bottom). Shown is the GIS's central
   $\sim 44$  arcmin  field   of  view   (images are  not   vignetting
   corrected). The pulsar is the   brightest source in the field.  Few
   faint point sources (labeled C1$-$C8)
   are detected in the neighborhood of
   Geminga.        }\label{gem_pwn_asca_gis2}\label{gem_pwn_asca_gis3}
   \end{figure*}

  In  order to  check the significance   of  this putative pulsar-wind
  nebula  and    to probe the pulsar   surrounding   in more   detail we
  reanalyzed the archival ASCA data of Geminga.  The pulsar was observed  
  by ASCA in   1994, March 28 for an   exposure  of about 71   ksec. The
  observation was performed with  an instrument setting similar  to that
  used  for the observation  of  PSR  B0656+14 (i.e.~$1\times 1$  arcmin
  pixel size for  the    GIS detectors,  source  at the   1-CCD  nominal
  position).  Unlike    for  PSR  B0656+14,  rise-time information   was
  recorded during  the observation so that  we  could apply the standard
  screening criteria including a data cleaning  for cosmic- and particle
  events. The  data of GIS2 and GIS3  where analyzed separately  for the
  known reasons   (see   Sect.~\ref{0656}). The  GIS2  and  GIS3  images,
  smoothed  with an adaptive  top-hat kernel (Ebeling  et al.~1999), are
  shown in Fig.\ref{gem_pwn_asca_gis2}. 

  In the   GIS2  image (Fig.\ref{gem_pwn_asca_gis2}a)  the  pulsar  is
  observed about 5 arcmin off-axis.  Few faint point sources (C1-C3)
  are     visible  close to  the   optical  axis  where the detector's
  sensitivity and  spatial resolution  is higher. Three  more sources
  (C4-C6) are visible in the western and north-eastern direction from the
  pulsar. 
  In   the  GIS3 image the pulsar   is observed on-axis  
  (Fig.\ref{gem_pwn_asca_gis2}b) and  the faint sources  C2$-$C4 are not
  detected due to the  decrease of instrument sensitivity off-axis.  Two
  additional but  rather faint sources  (C7$-$C8)  show up  in the  GIS3
  image. 

  Neither the GIS2 nor the GIS3 images show significant diffuse emission
  which can be  associated  with a pulsar-wind nebula.
  In particular, comparing Fig.2 of Kawai et al. (~1998b) with the images
  shown in Fig.\ref{gem_pwn_asca_gis2}a,b we find  that the shape of the  
  proposed pulsar-wind nebula follows  the pattern of the point  sources 
  C1$-$C6.  This coincidence leads us to conclude that the nebula is not 
  real but  is the result of an incomplete source subtraction. 

  Since no BeppoSAX observation  of Geminga  was  performed so far  we
  concentrate on the ROSAT data to search for counterparts of the ASCA
  clumps. ROSAT has observed the Geminga field in March-April 1991/92
  and  1996  with both   the PSPC and  HRI  detectors.  The effective
  exposure   of the merged   data sets,  which   we use  for our
  analysis, are $\sim 54$ ksec and $\sim 37$ ksec for the PSPC and HRI
  detectors, respectively.

  \begin{figure*}
  \centerline{\psfig{file=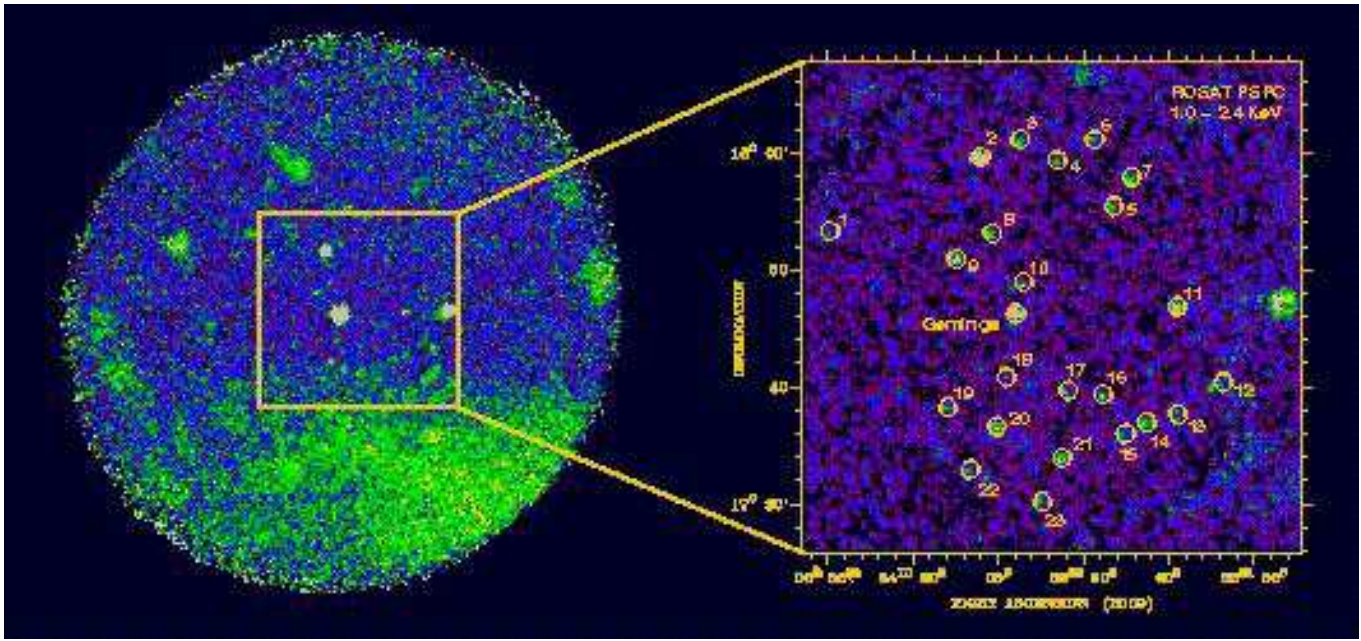,width=18cm,clip=}}
  \caption[h]{The  Geminga field  as seen  by  the  ROSAT PSPC in  the
  energy range  0.1-2.4 keV.  Shown is the  PSPC's $2^o$ full field of
  view (exposure and vignetting  corrected).  The pulsar is  the bright
  central point source. The diffuse and fuzzy emission which dominates
  the southern  part of the  detector field   belongs to  the  Monogem
  ring (Fig.\ref{monogem}).  The insert  marks the inner  region of  
  the PSPC detector. 23
  X-ray sources are detected beyond 1  keV in a $40 \times 40$ arcmin
  square centered on the pulsar.  Their positions are indicated in the
  hard band  image  shown on the   right. The  positions of   the PSPC
  sources  \#18,17,15,13,11,9,20 and  \#21  are in  agreement with the
  location  of  the X-ray clumps C1$-$C8   detected  in the  ASCA GIS
  data.\label{gem_pwn_rosat}} 
  \end{figure*} 

\begin{figure*}[th]
   \begin{picture}(160,135)(0,0)
   \put(0,0){\psfig{file=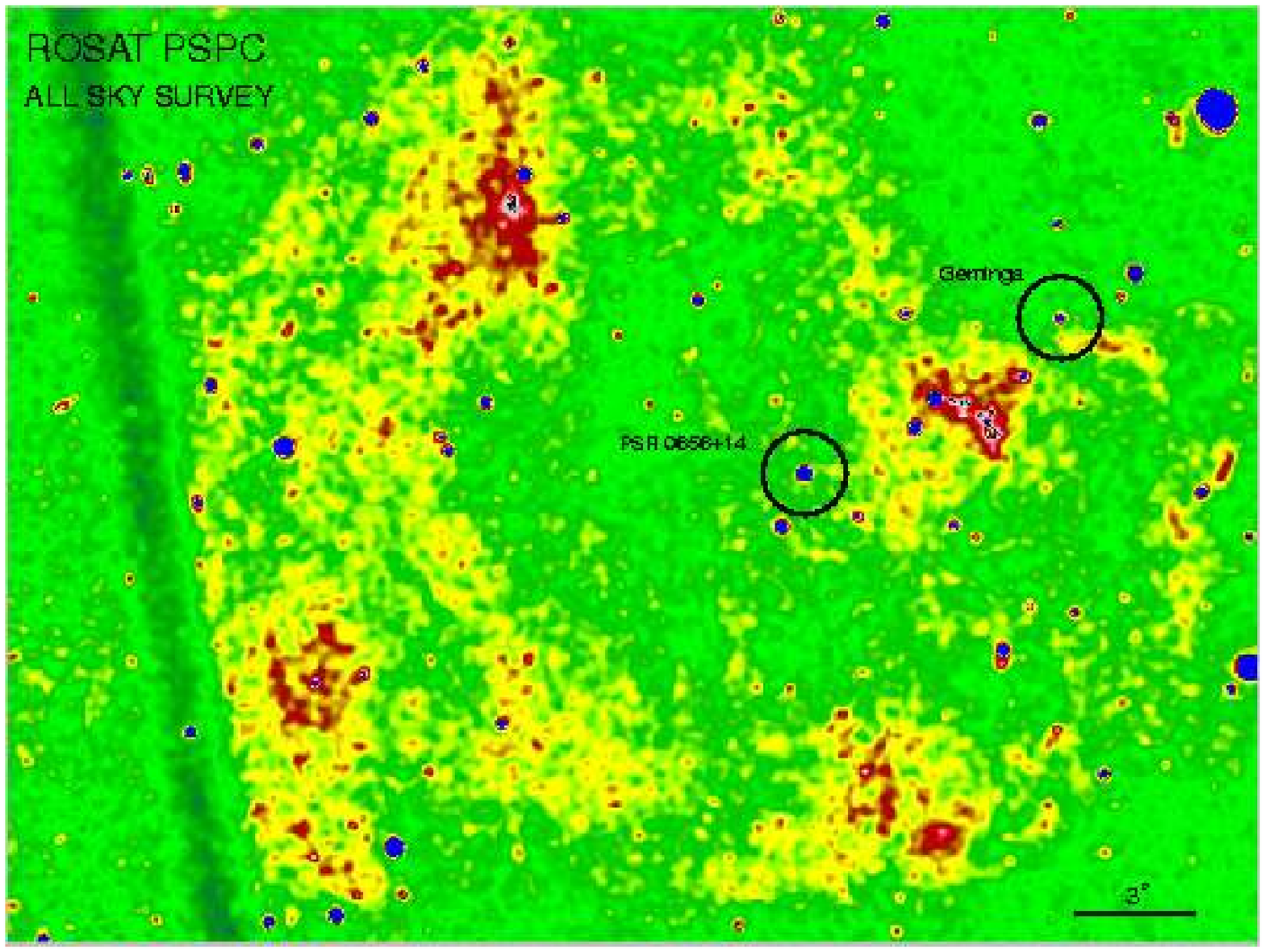,width=18cm,clip=}}
   \end{picture}
   \caption{\quad The $\sim 20^o$ wide Monogem ring as observed 
   in the ROSAT all-sky survey. The PSPC FOV during the pointed
   observations of Geminga and PSR 0656+14 are indicated. The 
   image demonstrates that both pulsars are located in crowded 
   regions with fuzzy background emission, difficult to disentangle
   with ASCA's spatial resolution of $\sim 3$ arcmin.\label{monogem}}
   \end{figure*}

  The PSPC image of the Geminga region showing the full field of view 
  as well as a zoomed hard band image is given in Fig.~\ref{gem_pwn_rosat}. 
  In order   to perform  a  spatial  analysis of the  pulsar  field we
  applied a Maximum-likelihood source detection in the soft (0.1$-$1.0
  keV)  and  in   the  hard    band   (1.0$-$2.4  keV)  using     {\em
  e\raisebox{-.22ex}{X}sas}. In total, 73  sources are detected with a
  significance of more  than 5  $\sigma$ in  the PSPC full  field of
  view. This implies a density of soft  X-ray sources which is about a
  factor of   two higher  than observed in   the neighborhood   of PSR
  0656+14 and PSR 1055$-$52. 
  The reason for that is that Geminga is  located close to the edge of
  the Monogem  ring (Fig.\ref{monogem}), a $\sim 20$ degree wide object 
  which is believed to  be an old and close by supernova       remnant
  (\cite{Plucinsky_96}). Actually,  a large  fraction of  the detected
  sources  is found to  be diffuse and  fuzzy emission of a
  small part of  the Monogem ring that  dominates the southern part of
  the detector field (see Fig.\ref{gem_pwn_rosat}). 
  The emission from the Monogem ring is very soft  so that in the hard
  band beyond 1 keV only 36 sources are detected in  the full field of
  view. 
  23 of  these sources are  located  at off-axis angles $\le  20$ arcmin,
  matching the GIS full field of view.   Their properties are summarized
  in Table \ref{gem_rosat_table}. 

  Inspecting the HRI data only  Geminga and the PSPC source \#2 are  
  detected. This  source was already  noticed in the Einstein  data as
  well      as        in    a short       ROSAT       HRI  observation
  (\cite{Becker_et_al93}1993). Its  improved HRI   position, extracted
  after   correcting the  data   for a 3   arcsec  attitude error
  recently found  to be  present in all  HRI  pointings,  is RA=06:34:06.6,
  DEC=17:59:53,  in  agreement with  its   PSPC  position. We  further
  checked the source extent of Geminga  and found it to be in agreement with
  the HRI point-spread function. 
  
  After a spatial correlation of the ROSAT and ASCA frames, we have
  identified the PSPC sources \#18,17,15,13,11,9,20 and \#21 as the
  likely counterparts of the hard X-ray clumps C1$-$C8. The faint diffuse and
  bridge-like emission  seen in the  GIS2  image south  of the sources
  C1$-$C3 is possibly due to the unresolved emission of the sources \#19$-$23. 

  A search for optical counterparts within a 10 arcsec error circle of
  the sources C1-C8 was  performed  using the  digitized POSS-II
  plates. Five of them were found to have optical candidate counterparts 
  (Table \ref{gem_rosat_table}). A correlation with the 1.4 GHz NRAO/VLA  
  Sky Survey (\cite{Condon_et_al98}1998) revealed radio counterparts for 
  the sources \#4 and \#9 (C6), using a search radius of 15 arcsec.

  \begin{table*}
   \caption[]{Properties of the ROSAT PSPC sources detected within a region 
    of $\le 20$ arcmin around Geminga. \label{gem_rosat_table}}
    \begin{flushleft}
    \begin{tabular}{c c c c c c c c c c}\hline\\[-1.5ex]
     Nr.     &   Range      &     Name     &  ASCA   & RA(2000)   &  DEC(2000) &       \Rate       &   mag   & mag   & ID   \\
     {}      &     keV      &      RX      &    ID   &   {}       &    {}      &        cts/s      &    B    &  R    & {}   \\[0.5ex]\hline\\[-1ex]
      1      &  1.0-2.4 & J0635.0+1753 &    {}   & 06 35 00.52 & 17 53 16.6  &   0.72 $\pm$ 0.17 &  {}  & {} &                 \\
      2      &  0.1-2.4 & J0634.1+1759 &    {}   & 06 34 06.92 & 17 59 45.9  &  34.47 $\pm$ 0.88 &  {}  & {} &                 \\
      3      &  0.1-2.4 & J0633.8+1801 &    {}   & 06 33 53.16 & 18 01 13.9  &   2.08 $\pm$ 0.32 &  {}  & {} &                 \\
      4      &  1.0-2.4 & J0633.6+1759 &    {}   & 06 33 39.45 & 17 59 28.9  &   0.45 $\pm$ 0.11 &  {}  & {} &                 \\
      5      &  0.1-2.4 & J0633.3+1755 &    {}   & 06 33 18.96 & 17 55 22.5  &   0.94 $\pm$ 0.21 &  {}  & {} &                 \\
      6      &  0.1-2.4 & J0633.4+1801 &    {}   & 06 33 26.04 & 18 01 07.5  &   1.37 $\pm$ 0.29 &  {}  & {} &                 \\
      7      &  0.1-2.4 & J0633.2+1757 &    {}   & 06 33 12.52 & 17 57 54.1  &   1.69 $\pm$ 0.29 &  {}  & {} &                 \\
      8      &  0.1-2.4 & J0634.0+1753 &    {}   & 06 34 02.89 & 17 53 05.3  &   1.10 $\pm$ 0.20 &  {}  & {} &                 \\
      9      &  0.1-2.4 & J0634.2+1750 &    C6   & 06 34 15.30 & 17 50 58.9  &   2.46 $\pm$ 0.26 & $>$22.5 & $>$20.8  &        \\
     10      &  0.1-2.4 & J0633.8+1748 &    {}   & 06 33 51.47 & 17 48 44.6  &   1.67 $\pm$ 0.24 &  {}  & {} &                 \\
     Gem     &  0.1-2.4 & J0633.9+1746 &    {}   & 06 33 54.33 & 17 46 11.9  &  504.24$\pm$ 3.10 &  {}  & {} &                 \\
     11      &  0.1-2.4 & J0632.9+1746 &    C5   & 06 32 56.35 & 17 46 43.9  &   3.41 $\pm$ 0.32 & $>$22.5 & $>$ 20.8    & {}  \\
     12      &  1.0-2.4 & J0632.6+1740 &    {}   & 06 32 39.64 & 17 40 07.7  &   1.01 $\pm$ 0.20 &  {}  & {} &                 \\
     13      &  0.1-2.4 & J0632.9+1737 &    C4   & 06 32 55.97 & 17 37 34.1  &   1.58 $\pm$ 0.32 &  18.9  & 18.3  & nS         \\
     14      &  0.1-2.4 & J0633.1+1736 &    {}   & 06 33 06.50 & 17 36 41.8  &   1.43 $\pm$ 0.27 &  {}  & {} &                 \\
     15      &  1.0-2.4 & J0633.2+1735 &    C3   & 06 33 14.99 & 17 35 41.4  &   0.94 $\pm$ 0.18 & $>$22.5  & $>$20.8    & {}  \\
     16      &  1.0-2.4 & J0633.3+1739 &    {}   & 06 33 22.99 & 17 39 05.1  &   0.61 $\pm$ 0.12 &  {}  &   &                  \\
     17      &  1.0-2.4 & J0633.5+1739 &    C2   & 06 33 35.53 & 17 39 27.2  &   0.33 $\pm$ 0.09 &  14.3  & 12.7  &  ?         \\
     18      &  0.1-2.4 & J0633.9+1740 &    C1   & 06 33 56.89 & 17 40 46.0  &   1.26 $\pm$ 0.22  &  16.9  & 17.5  &  ?        \\
     19      &  1.0-2.4 & J0634.3+1737 &    {}   & 06 34 18.52 & 17 37 58.2  &   0.33 $\pm$ 0.09 &  {}  & {} &                 \\
     20      &  0.1-2.4 & J0634.0+1736 &    {}   & 06 34 01.47 & 17 36 15.7  &   2.45 $\pm$ 0.30 &  {}  & {} &                 \\
     21      &  0.1-2.4 & J0633.6+1733 &    C7   & 06 33 37.41 & 17 33 41.4  &   1.75 $\pm$ 0.26 &  18.4  & 21.0  & nS         \\
     22      &  0.1-2.4 & J0634.2+1732 &    C8   & 06 34 10.53 & 17 32 30.0  &   1.23 $\pm$ 0.25 &  $>$22.5  & $>$ 20.8    & {}   \\
     23      &  1.0-2.4 & J0633.7+1729 &    {}   & 06 33 44.78 & 17 29 48.8  &   0.55 $\pm$ 0.13 &    &   &    \\[1ex]\hline\\[-4ex]
  \end{tabular}
  \end{flushleft}
  $^a$background, dead-time and vignetting corrected rate in the given energy range.
  \end{table*}

\section{Summary and discussion}

  Based on ASCA observations of rotation-powered pulsars Kawai \& Tamura (1996),
  \cite{Shibata_et_al97}(1997) and \cite{Kawai_et_al98}(1998a, 1998b)  
  have  reported on the possible existence of huge, $10-20$  arcmin wide 
  X-ray bright nebulae  around most of the ASCA  detected pulsars.  The  
  existence of these nebulae was explained by an  interaction of a  
  pulsar-wind outflow  with the surrounding  interstellar   matter.    
  In   particular,   the  X-ray luminosity  of these nebulae was found  
  to correlate with the pulsar spin-down energy  $\dot{E}$ 
  (\cite{Kawai_et_al98}1998b). Interpreted  as   a universal phenomenon, 
  \cite{Shibata98}(1998) used these  pulsar-wind nebulae as a calorimeter 
  to constrain  parameters of the pulsar wind and the confining pressure 
  of the ambient medium. 

  In order to asses the reality of these nebulae, we have performed a
  careful study  of the  fields    around the three  musketeers   PSR
  B1055$-$52, B0656+14, and Geminga, based on a combined analysis
  using archival ASCA, BeppoSAX and ROSAT data.
  
  Based on these data we could not establish the reality of the $10-20$ 
  arcmin wide X-ray bright and clumpy pulsar-wind  nebula  around the 
  three musketeers. In all cases faint, point-like, X-ray sources were  
  found and clearly resolved by the ROSAT PSPC, well separated from  the  
  pulsar and  coincident with the knots reported by \cite{Shibata_et_al97}(1997) 
  and \cite{Kawai_et_al98}(1998a, 1998b). On a larger scale, our analysis 
  showed that the spatial distribution of these sources follows quite well 
  the shape of the putative X-ray nebulae. For most of them we have found 
  candidate optical counterparts in digitized photographic plates and for 
  the sources around PSR B1055$-$52 we also identified likely radio 
  counterparts in data taken by \cite{Stapper_et_al99}(1999). NVSS radio
  data (Condon et al.~1998) were used to search for radio counterparts of
  the X-ray sources around Geminga and PSR 0656+14.

  Although the visual appearance of weak emission patterns can sensitively 
  depend on details of the image construction method, the existence of 
  optical and radio counterparts for many of these X-ray sources strongly 
  favors their interpretation as isolated point sources.
  We therefore conclude that the apparent clumpy extended  emission
  observed in ASCA data was probably caused by the unresolved contribution 
  of point sources located in the vicinity of the pulsars, enhanced  by 
  data-analysis artifacts like incomplete source subtraction, background 
  modeling and Gaussian smoothing.

  We remind that the three musketeers are located  close to the galactic
  plane, i.e.~in crowded regions.   Indeed, our analysis has shown that,
  on average, about 20 hard X-ray sources (e.g. identified beyond 1 keV)
  are detected with  the PSPC in a  region which corresponds to  the GIS
  full field of view. Even if most of these sources  are not detected by
  the GIS,  they contribute to  the  background emission which --  by the
  lack of spectral information -- is difficult to model.

  These results are in line with similar findings recently reported by
  \cite{Brinkman_et_al99}(1999)    and  \cite{Pivovaroff_et_al99}(1999)
  who could not confirm the existence of the putative pulsar-wind nebula
  surrounding PSR B1610$-$50 and PSR B1046$-$58.

  Of course, the sensitivity of our pulsar-wind nebulae search is background
  limited. However, given the lack of models predicting the extent or
  the surface brightness of the nebula as a function of the pulsars'
  parameters we restrain from giving explicit upper limits as this would
  require {\em ad hoc} assumptions on a nebula size and properties. The 
  different behavior observed for the plerions around the Crab-, PSR 1509-58 
  and Vela support this view. While for the Crab-nebula (d=2 kpc, extent 
  $\sim 2'$) the emission efficiency in terms of the pulsar spin-down 
  energy  is $\sim 5\%$ in the ROSAT band, it is only $\sim 1\%$ for the 
  nebula around PSR 1509-58 (d=4.9 kpc, extent $7'\times 9'$) and 
  $\sim 0.04\%$ for the Vela-nebula (d=0.5 kpc, extent 2') 
  (Becker \& Tr\"umper 1997).
  Finally, we note that the results of our analysis does not argue against
  the presence of compact synchrotron nebulae on smaller scales.  However,
  the ROSAT HRI data available for all of the investigated sources restrict
  their extension to less than $\sim 10$ arcsec. Providing a resolving power
  of 1 arcsec it will be interesting  to  see what information Chandra will
  add to this issue.

\begin{acknowledgements}

  The ROSAT project is supported by the Bundesministerium f\"ur Bildung,
  Wissenschaft, Forschung und Technologie (BMBW) and the Max-Planck-Society
  (MPG). We thank our colleagues from the MPE ROSAT group for their support.
  WB \& WB thank the Cosmic Radiation Laboratory for hospitality where part
  of the research was done in the framework of the RIKEN-MPE exchange program.
  WB is grateful to G.~Cusumano from IFCAI/CNR for support during the reduction
  of BeppoSAX data. On behalf of the GSC2 Project collaboration, RM wants to
  dedicate his contribution to the present paper to the memory of Barry Lasker.
  The Guide Star Catalogue-II is produced by the Space Telescope Science
  Institute, in collaboration with the Osservatorio Astronomico di Torino.
  Additional support is provided by the Association of Universities for Research
  in Astronomy, the Italian Council for Research in Astronomy, European Southern
  Observatory, Space telescope European Coordinating facility, the International
  GEMINI project and the European Space Agency Astrophysics Division.

\end{acknowledgements}

\end{document}